\documentclass[11pt,preprint,aps,showpacs]{revtex4-1}
\usepackage{mathrsfs}
\usepackage{amsmath}
\usepackage{amssymb}
\usepackage{epsfig}
\usepackage{graphicx}
\usepackage{booktabs}
\usepackage{array}
\usepackage{multirow}
\usepackage{slashed}
\usepackage{color}
\usepackage{braket}
\usepackage[colorlinks=true,linkcolor=red,citecolor=blue]{hyperref}
\allowdisplaybreaks[4]
\begin{document}
\renewcommand{\arraystretch}{0.5}
\newcommand{\beq}{\begin{eqnarray}}
\newcommand{\eeq}{\end{eqnarray}}
\newcommand{\non}{\nonumber\\ }
\newcommand{\acp}{ {\cal A}_{CP} }
\newcommand{\psl}{ p \hspace{-1.8truemm}/ }
\newcommand{\nsl}{ n \hspace{-2.2truemm}/ }
\newcommand{\vsl}{ v \hspace{-2.2truemm}/ }
\newcommand{\epsl}{\epsilon \hspace{-1.8truemm}/\,  }
\newcommand{\tf}{\textbf}
  \def\tb{\textcolor{blue}}
  \def\tr{\textcolor{red}}
\def \cpl{ Chin. Phys. Lett.  }
\def \ctp{ Commun. Theor. Phys.  }
\def \epjc{ Eur. Phys. J. C }
\def \jpg{  J. Phys. G }
\def \npb{  Nucl. Phys. B }
\def \plb{  Phys. Lett. B }
\def \prd{  Phys. Rev. D }
\def \prl{  Phys. Rev. Lett.  }
\def \zpc{  Z. Phys. C }
\def \jhep{ J. High Energy Phys.  }

\title{Study of the $B^{+}\to\pi^{+}\left(\pi^{+}\pi^{-}\right)$ decay in PQCD Approach}
\author{
Qin Chang$^{1,3}$ \footnote{changqin@htu.edu.cn},
Lei Yang$^1$ \footnote{yanglei@stu.htu.edu.cn},
Zhi-Tian Zou$^2$ \footnote{zouzt@ytu.edu.cn},
Ying Li$^2$ \footnote{liying@ytu.edu.cn}
}
\affiliation
{\small
1.~Institute of Particle and Nuclear Physics,Henan Normal University, Xinxiang 453007, China\\
2.~Department of Physics, Yantai University, Yantai 264005, China\\
3.~Institute of Physics, Henan Academy of Sciences, Zhengzhou 455004, China
}
\begin{abstract}
Based on the fitting results of the LHCb collaboration on the contributions of various intermediate resonances to the $B^{+}\to\pi^{+}\pi^{+}\pi^{-}$ decay, we make systematically calculate the branching fractions and localized $CP$ asymmetries of the quasi-two-body $B^{+}\to \pi^{+} \left( \rho(770), \omega(782), \rho(1450), f_{2}\left(1270\right)\to\right)\pi^{+}\pi^{-}$ decays within the perturbative QCD (PQCD) approach. Our theoretical predictions for the branching fractions agree well with the data within the errors. In order to further test the framework of the three-body $B$ decays in PQCD and the wave functions of $\pi\pi$ pair, we also calculate the branching fractions of the corresponding two-body $B^{+}\to\pi^{+}\rho(770)$, $B^{+}\to\pi^{+}\omega(782)$ and $B^{+}\to\pi^{+}f_{2}\left(1270\right)$ decays under the narrow-width-approximation, which are in consistence with the experimental data. The direct $CP$ asymmetries of  $B^{+}\to\pi^{+}(\rho(770/1450)\to) \pi^{+}\pi^{-}$ decays are found to be very small because these decay modes are tree-dominated. However, due to the large penguin contributions from the chiral enhanced annihilation diagrams, $B^{+}\to\pi^{+} (f_{2}\left(1270\right)\to )\pi^{+}\pi^{-}$ decay has a large direct $CP$ asymmetry, which is also discovered by LHCb collaboration. A relatively large  $CP$ asymmetry  is also expected in the $B^+\to\pi^+ (\omega\to)\pi^+\pi^-$ decay occurring via $\rho-\omega$ mixing, which would be measured LHCb  and Belle-II experiments in future.
\end{abstract}
\maketitle
\section{Introduction}
The three-body $B$ meson decays sharing a large fraction of the branching ratio play important roles in deeply testing the standard model~(SM) and  understanding the mechanism of the $CP$ violation especially the sources of the strong phases. Experimentally, a lot of three-body $B$ meson decays have been observed by the BaBar \cite{BaBar:2003zav, BaBar:2005qms, BaBar:2006hyf, BaBar:2007aut, BaBar:2007hmp, BaBar:2009jov, BaBar:2009vfr}, Belle \cite{Belle:2002ryx, Belle:2003ibz, Belle:2004drb, Belle:2005rpz, Belle:2006ljg, Belle:2008til, Belle:2010wis}, and LHCb \cite{LHCb:2013lcl, LHCb:2014mir, LHCb:2016vqn, LHCb:2017hbp, LHCb:2018oeb, LHCb:2019sus,LHCb:2019vww} collaborations. With the steady accumulation of experimental data, the theoretical researches for the three-body $B$ meson decays have been studied extensively in the approach based on the symmetry principles \cite{Gronau:2005ax, Engelhard:2005hu, He:2014xha},  for instance, the QCD factorization (QCDF) \cite{El-Bennich:2009gqk, Cheng:2002qu, Cheng:2007si, Cheng:2013dua, Li:2014oca, Chen:2023pms}, the perturbative QCD approach (PQCD) \cite{Yan:2023yvx, Li:2018psm, Zou:2020ool, Chen:2002th, Li:2018lbd, Li:2016tpn, Zou:2022xrr, Liu:2021sdw, Yao:2022zom, Yang:2021zcx, Zou:2020mul, Zou:2020fax, Li:2021cnd}, and other theoretical methods \cite{Wang:2015ula, El-Bennich:2006rcn,Chang:2019obq,Guo:2000uc,Cheng:2016shb}. In comparison with the two-body $B$ meson decays, the dynamic of the three-body $B$ meson decay is much more complicated  because the momentum of each final particle is a variable. In addition, the three-body decays include both resonant and non-resonant contributions, and how to exactly distinguish and evaluate these contributions is important for studying the three-body $B$ meson decays.

In analyzing three-body $B$ meson decays, the Dalitz plot is a widely used and effective method, which can be divided into different regions in term of the characteristic kinematics. The central region in the Dalitz plot corresponds to the nonresonant contributions, which are $\alpha_s$ suppressed relative to the contributions of edges \cite{Virto:2016fbw}. The edges of the Dalitz plot correspond to the kinematical configuration  where  two of  final particles  fly collinearly, generating an invariant mass recoiling against the third bachelor meson. In this region, the two collinear mesons move almost in the same direction  and the bachelor recoils back.  The two collinear mesons  can be viewed as a cluster, and might form resonances with different angular momenta. Theoretically, we can refer to the three-body B meson decays in this region as the quasi-two-body decays beyond the narrow-width approximation \cite{Virto:2016fbw}, where the interaction in the two-meson pair can be described by the nonperturbative two-meson wavefunction \cite{Hua:2020usv, Li:2006cva, Yan:2023yvx, Li:2018psm, Zou:2020ool}. Therefore, we can  evaluate these quasi-two-body decays theoretically within the QCD-inspired approaches such as PQCD and QCDF. Experimentally, the resonance contributions are analyzed by using some methods, such as the isobar model \cite{Herndon:1973yn, Sternheimer:1961zz}, the K-matrix formalism \cite{Chung:1995dx} and the quasi-model independent analysis\cite{LHCb:2019sus},  among which the isobar model method is widely adopted by BaBar, Belle and LHCb collaborations. 

In the so-called isobar model, the decay amplitude can be modeled as a coherent sum  of the amplitudes of  ``$N$" individual decay channels corresponding to different resonances with same spin. The decay amplitude can be written as
\begin{align}
\mathcal{A}=\sum_i^N a_i \mathcal{A}_i\,,
\label{isb}
\end{align}
where $\mathcal{A}_i$ is the decay amplitude of the channel  associated with a certain resonance and $a_i$ is the corresponding effective coefficient reflecting the relative magnitude and phase. From eq.~(\ref{isb}),  it can be easily found that not only does the individual $\mathcal{A}_i$ provide the  possible source of $CP$ asymmetry, the interference between them also lead to $CP$ asymmetry.  The effective coefficient $a_i$ in eq.~(\ref{isb}) can only be determined from the experimental data now, while the amplitude, $\mathcal{A}_i$, can be calculated perturbatively by the QCD-inspired approaches such as PQCD, which will be detailed in the next section.

In Ref.~\cite{LHCb:2019sus}, based on the data sample with an integrated luminosity of 3 $fb^{-1}$ of $pp$ collisions, LHCb collaboration reported the results of an amplitude analysis of the three-body decay $B^+\to\pi^+\pi^+\pi^-$, where the $CP$ asymmetry effects are also taken into account. The nonresonant component is described by the behavior of the $S$-wave contribution, The resonant contributions are described by using the isobar model, and include  $\rho(770)$, $\omega(782)$ and $\rho(1450)$ resonances in the $\pi\pi$ $P$-wave, the $f_2(1270)$ resonance in the $\pi\pi$ $D$-wave and the $\rho_3(1690)$ resonance in the $\pi\pi$ $F$-wave. In the $B^+\to \pi^+(f_2(1270)\to)\pi^+\pi^-$ process, large $CP$ asymmetry was measured. Further phenomenological  and experimental investigations are required to shed light on the underlying  mechanism of the three-body decays and searching for new $CP$ asymmetry effects in the three-body decays. In Ref.~\cite{Li:2016tpn},  the contribution of $P$-wave state $\rho(770)$ to $B^+\to \pi^+\pi^+\pi^-$ decay has been studied in PQCD approach, while the possible effect of $\rho-\omega$ mixing has not been considered. The contribution of $D$-wave $f_2(1270)$ resonance state was also studied in  Ref.~\cite{Li:2018lbd}. For completeness, we will investigate above-mentioned $P$-wave and $D$-wave resonant contributions to the $B^+\to \pi^+\pi^+\pi^-$ decay simultaneously in this work and the interference effect between them will be studied. In additional, the resonant contribution of $\omega(782)$ will be induced due to the $\rho-\omega$ mixing.

This paper is organized as follows. In  section II, the theoretical framework of PQCD approach for the quasi-two-body decay  is reviewed, and the $P$ wave and $D$ wave two-pion wavefunctions are discussed  briefly. In section III, the input parameters are collected, our numerical results and some discussions are given in detail. Finally, we  conclude with a summary in section IV. The explicit  expressions of decay amplitudes within the PQCD approach are collected in appendix A.

\section{Theoretical Framework}\label{sec:function}

The PQCD approach is based on $k_T$ factorization, where the small intrinsic transverse momenta of quarks are kept in the calculation. As a result, the end-point singularity is avoided,  while the double logarithm will appear in the QCD radiative corrections caused by the additional scale introduced by the transverse momentum. Fortunately, these double logarithm can be resumed within the renormalization group equation, leading to the so-called Sudakov form factor that suppresses the long distance contributions effectively and make the perturbative calculation reliable. Moreover, within this framework the annihilation diagrams can be calculated perturbatively, and provide the dominant source of the strong phase, which is required by the direct $CP$ asymmetry in SM \cite{Hong:2005wj}.

For the $B$ meson decay involving multi scales, the physics can be decomposed in terms of the characteristic scales.  It is known that there are  three characteristic energy scales in the $B$ meson decays including the electroweak scale $m_W$,  the scale of the B-decays $m_b$ and the factorization scale $\sqrt{\overline{\Lambda}m_B}$ with $\overline{\Lambda}=m_B-m_b$. The Wilson coefficients summarize the short-distance physical contributions, and are in principle computable with the perturbative theory order by order at the scale of $\mu\sim m_W$. Renormalization group equation enable us to evaluate the dynamical effects in scaling the Wilson coefficients from $m_W$ to $m_b$ scale. The physics between the $m_b$ and the factorization scale is involved in the hard kernel $\mathcal{H}$, which is dominated by the hard gluon exchanging process, and can be calculated perturbatively in PQCD approach. The physics below the factorization scale is soft and nonperturbative, and can be described by the wave function.  Therefore,  in the framework of PQCD, the amplitude of the quasi-two-body $B$ meson decays $B \to \pi \left(R\to \right)\pi \pi $ can be decomposed into the convolution as\cite{Chen:2002th}
\begin{eqnarray}
	\mathcal{A}=C\left(t\right)\otimes\mathcal{H}\left(x_{i},b_{i},t\right)\otimes\Phi_{B}\left(x_{1},b_{1}\right)\otimes\Phi_{\pi}\left(x_{2},b_{2}\right)\otimes\Phi_{\pi\pi}\left(x_{3},b_{3}\right)\otimes \exp[-S\left(t\right)],
\end{eqnarray}
where $x_{i}$ are the momentum fractions of  quarks, $b_{i}$ are the conjugate variables of the transverse momentum of quarks, and $t$ is the largest scale that appears in the hard kernel $\mathcal{H}$. $\Phi_{B}$ and $\Phi_{\pi}$ are the wave functions of the $B$ and $\pi$ mesons, respectively, and $\Phi_{\pi\pi}$ is the two-pion  wave function, which involves the resonant and the non-resonant interactions that between the two collinear $\pi$ mesons. The exponential term is the Sudakov form factor, which can suppress the soft dynamic effects \cite{Li:2012md, Li:2012nk, Li:2001ay, Li:1997un}.

\begin{figure}[t]
	\centering
	\includegraphics[width=0.7\linewidth]{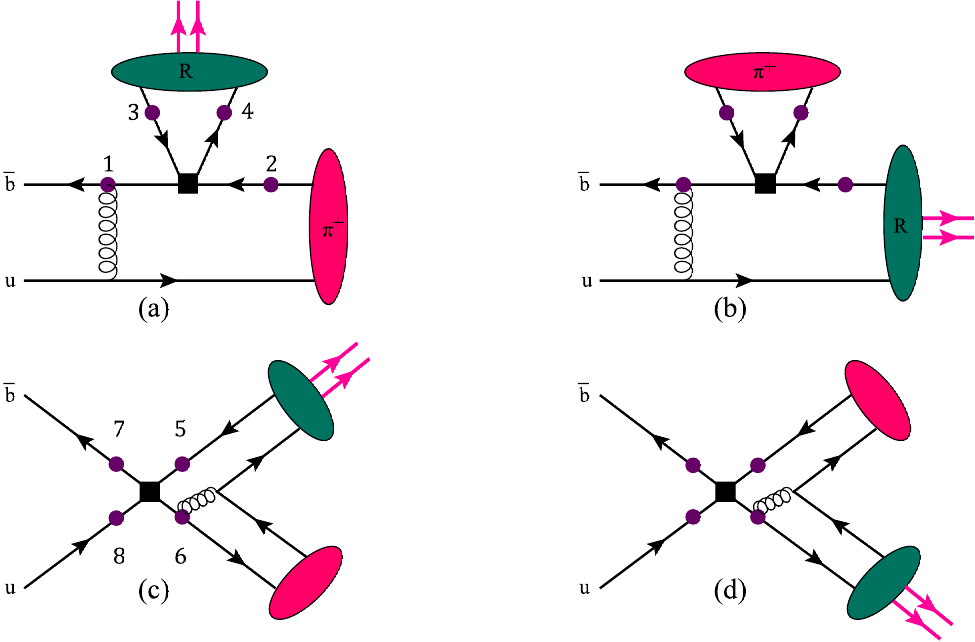}
	\caption {The leading  order Feynman diagrams contributing to the
		$B^{+}\to\pi^{+}\left(R\to\right)\pi^{+}\pi^{-}$ ($R=\rho(770), \omega(782), \rho(1450), f_2(1270)$) decays in PQCD approach, where the black squares stand for the weak vertices,  the purple  spots on the quark lines denote possible attachments of hard  gluons, the green ellipses represent $\pi^{+}\pi^{-}$-pair and the red ones are the light bachelor $\pi^{+}$ mesons.}
	\label{fig:diagram1}
\end{figure}

The effective weak Hamiltonian responsible for  the $b \to q\bar{q} d$ transition is written as
\begin{eqnarray}
	\mathcal{H}_{eff} =\frac{G_{F}}{\sqrt{2}}  \sum_{p=u,c} V_{pb}V^{\ast}_{pd} \left[  C_{1}\left(\mu\right)
	O_{1}^p\left( \mu\right)+ C_{2}\left(\mu\right)O_{2}^p\left(\mu\right) +
	\sum_{i=3}^{10}C_{i}\left(\mu\right)O_{i}\left(\mu\right) \right]
\end{eqnarray}
with the current-current operators:
\begin{eqnarray*}
		O_{1}\left(\mu\right)=\left(\overline{p}_{\alpha} b_{\beta}\right)_{V-A}\left(\overline{d}_{\beta}
		p_{\alpha}\right)_{V-A},\ \ \
		O_{2}\left(\mu\right)=\left(\overline{p}_{\alpha}b_{\alpha}\right)_{V-A}\left(\overline{d}_{\beta}
		p_{\beta}\right)_{V-A},
\end{eqnarray*}
the QCD penguin operators:
\begin{eqnarray*}
		O_{3}\left(\mu\right)&=&\left(\overline{d}_{\alpha} b_{\alpha}\right)_{V-A}\sum_{q }^{}\left(\overline{q} _{\beta}
		q _{\beta}\right)_{V-A},\ \ \
		O_{4}\left(\mu\right)=\left(\overline{d}_{\beta} b_{\alpha}\right)_{V-A}\sum_{q }^{}\left(\overline{q} _{\alpha}
		q _{\beta}\right)_{V-A},\\
		O_{5}\left(\mu\right)&=&\left(\overline{d}_{\alpha} b_{\alpha}\right)_{V-A}\sum_{q }^{}\left(\overline{q} _{\beta}
		q _{\beta}\right)_{V+A},\ \ \
		O_{6}\left(\mu\right)=\left(\overline{d}_{\beta} b_{\alpha}\right)_{V-A}\sum_{q }^{}\left(\overline{q} _{\alpha}
		q _{\beta}\right)_{V+A},
		\end{eqnarray*}
the electroweak penguin operators:	
	\begin{eqnarray*}
		O_{7}\left(\mu\right)&=&\frac{3}{2}\left(\overline{d}_{\alpha} b_{\alpha}\right)_{V-A}\sum_{q }^{}e_{q }\left(\overline{q} _{\beta}
		q _{\beta}\right)_{V+A},\ \ \ \
		O_{8}\left(\mu\right)=\frac{3}{2}\left(\overline{d}_{\beta} b_{\alpha}\right)_{V-A}\sum_{q }^{}e_{q }\left(\overline{q} _{\alpha}
		q _{\beta}\right)_{V+A},\\
		O_{9}\left(\mu\right)&=&\frac{3}{2}\left(\overline{d}_{\alpha} b_{\alpha}\right)_{V-A}\sum_{q }^{}e_{q }\left(\overline{q} _{\beta}
		q _{\beta}\right)_{V-A},\ \ \
		O_{10}\left(\mu\right)=\frac{3}{2}\left(\overline{d}_{\beta} b_{\alpha}\right)_{V-A}\sum_{q }^{}e_{q }\left(\overline{q}_{\alpha}
		q_{\beta}\right)_{V-A}, 				
\end{eqnarray*}
where $\left(\overline{q}_{1} q_{2}\right)_{V\pm A}=\overline{q}_{1}\gamma^{\mu}(1 \pm \gamma_5) q_{2}$, $\alpha$ and $\beta$ are  the color  indices,  $V_{pb}V^{\ast}_{pd}$ are the products of the CKM element,  $C_i(\mu)$ are the Wilson coefficients, and a summation over $q=u,d,s,c,b$ is implied.  With the effective weak Hamiltonian, the  leading order Feynman diagrams for the quasi-two-body decays $B^{+}\to\pi^{+}\left(R\to\right)\pi^{+}\pi^{-}$ ($R=\rho(770), \omega(782), \rho(1450), f_2(1270)$) are illustrated in Fig.~\ref{fig:diagram1}.

We consider the $B$ meson at rest for simplicity. In the light-cone coordinate, the momenta of $B$ meson and  light  quark in it can be written as
\begin{eqnarray}
	p_{B}=\frac{M_{B}}{\sqrt{2}}\left(1,1,0_{T}\right),\ \ k_{1}=\left(0,\frac{M_{B}}{\sqrt{2}}x_{1},k_{1T}\right),
\end{eqnarray}
with $x_{1}$ being the momentum fraction of light quark. The momenta of the $\pi\pi$ pair moving along $z$ direction and the recoiled bachelor $\pi$ meson are respectively written  as
\begin{eqnarray}
   p_{2}=\dfrac{M_{B}}{\sqrt{2}}\left(1,\eta,0_{T}\right), \quad p_{3}=\dfrac{M_{B}}{\sqrt{2}}\left(0,1-\eta,0_{T}\right),
\end{eqnarray}
where $\eta=s/M_{B}^{2}$,  and $\sqrt{s}$ is the invariant mass of the $\pi\pi$ pair satisfying $s=p^{2}$. The momenta of quarks in the $\pi\pi$ pair~(resonances) and the recoiled $\pi$ meson can be written  as
\begin{eqnarray}
k=\left(\dfrac{M_{B}}{\sqrt{2}}z,0,0_{T}\right), \quad k_{3}=\left(0,\left(1-\eta\right)x_{3},0_{T}\right),
\end{eqnarray}
where $z$ and $x_{3}$ are the momentum fractions of quark in the $\pi\pi$ pair and the recoiled $\pi$ meson, respectively. Besides,  the momenta of two $\pi$  in the $\pi\pi$ pair have the components
\begin{eqnarray}
	p_{1}^{+}=\zeta\dfrac{M_{B}}{\sqrt{2}},\ \  p_{1}^{-}=\left(1-\zeta\right)\dfrac{M_{B}}{\sqrt{2}},\ \ p_{2}^{+}=\left(1-\zeta\right)\dfrac{M_{B}}{\sqrt{2}},\ \ p_{2}^{-}=\zeta\dfrac{M_{B}}{\sqrt{2}}\,,
\end{eqnarray}
where $\zeta$ represents the momentum fraction of one pion in the $\pi\pi$ pair.


The wave functions of the initial and final states are the most important inputs in calculating the $B$ meson decays within the PQCD approach.  For $B$ meson and  $\pi$ meson, their wave functions have been widely studied and well defined in the non-leptonic two-body $B$ decays. In our following calculation, the ones given in Ref. \cite{Li:2003yj, Wang:2019msf, Rui:2021kbn}  are used. For the  $\pi\pi$ pair, the wave functions associated with the $P$- and $D$-wave resonances are relatively less studied. Because the initial state $B$ and the final state $\pi$ are pseudoscalar mesons,  only longitudinal polarization state the $\pi\pi$ pair contributes to the decay amplitude due to the conservation of angular momentum. The $P$-wave wave function of the longitudinal polarized two-pion pair is given as \cite{Rui:2021kbn, Li:2003yj, Wang:2019msf}
\begin{eqnarray}
   \Phi_{P}^{L}\left(z,\zeta,s\right)=\dfrac{1}{\sqrt{2N_{C}}}\left[\omega\slashed{\epsilon}_{P}\phi_{P}^{0}\left(z,\zeta,s\right)+\omega\phi_{P}^{s}\left(z,\zeta,s\right)+\frac{\slashed{p}_{1}\slashed{p}_{2}-\slashed{p}_{2}\slashed{p}_{1}}{\omega\left(2\zeta-1\right)}\phi_{P}^{t}\left(z,\zeta,s\right)\right].
\end{eqnarray}
where $\phi_{P}^{0}$ and $\phi_{P}^{s,t}$ are the twist-2 and  twist-3 distribution amplitudes (DAs), respectively, and can be written as \cite{Rui:2021kbn}
\begin{eqnarray}
	\phi_{P}^{0}\left(z,\zeta,s\right)&=&\frac{3F_{P}^{\parallel}\left(s\right)}{\sqrt{2N_{C}}}z\left(1-z\right)\left[1+a_{p}^{0}C_{2}^{3/2}\left(1-2z\right)\right]P_{1}\left(2\zeta-1\right),\nonumber \\
	\phi_{P}^{s}\left(z,\zeta,s\right)&=&\frac{3F_{P}^{\perp}\left(s\right)}{2\sqrt{2N_{C}}}z\left(1-2z\right)\left[1+a_{p}^{s}\left(1-10z+10z^{2}\right)\right]P_{1}\left(2\zeta-1\right),\nonumber \\
	\phi_{P}^{t}\left(z,\zeta,s\right)&=&\frac{3F_{P}^{\perp}\left(s\right)}{2\sqrt{2N_{C}}}z\left(1-2z\right)\left[1+a_{p}^{t}C_{2}^{3/2}\left(1-2z\right)\right]P_{1}\left(2\zeta-1\right),
\label{eq:DA1}
\end{eqnarray}
with the Legendre polynomial $P_{1}\left(2\zeta-1\right)=2\zeta-1$ and the Gegenbauer polynomial $C_{2}^{3/2}\left(t\right)=3\left(5t^{2}-1\right)/2$. The Gegenbauer moments $a_{p}^{0,s,t}$ in Eq.~\eqref{eq:DA1}  are left to be determined by combining experimental data.  For the time-like form factor $F_{P}^{\parallel}(s)$, we employee the  vector-dominance model. Then,   including the $\rho-\omega$ interference, it can be parameterized as Refs.~\cite{LHCb:2019sus, Wang:2016rlo, Li:2017obb, BaBar:2012bdw},
\begin{eqnarray}
	F_{P}^{\parallel}\left(s\right)&=&\left[GS_{\rho}\left(s,m_{\rho},\Gamma_{\rho}\right)\frac{1+c_{\omega}RBW_{\omega}\left(s,m_{\omega},\Gamma_{\omega}\right)}{1+c_{\omega}}+c_{\rho(1450)}RBW_{i}\left(s,m_{\rho(1450)},\Gamma_{\rho(1450)}\right)\right]\nonumber \\
&\times&	\frac{1}{1+c_{\rho(1450)}} \,, 
	\label{lineshape}
\end{eqnarray}
where $s=m^2(\pi\pi)$ is the two-pion invariant mass squared, $m_{\rho,\omega,\rho(1450)}$ and $\Gamma_{\rho,\omega,\rho(1450)}$ are the masses and decay widths of the corresponding resonances, respectively. $c_{\omega,\rho(1450)}$ are the weight factors relative to the $\rho\left(770\right)$ resonance. For the excited-states, only $\rho\left(1450\right)$ contribution is taken into account. The functions $GS_{\rho}$, $RBW_{\omega}$ and $RBW_{\rho(1450)}$ describe the mass lineshap of each resonance state respectively. 

Based on the analysis of $B^{+}\to\pi^{+}\pi^{-}\pi^{+}$ decay  in Ref.~\cite{LHCb:2019sus}, the central values of the Gegenbauer moments $a_{p}^{0,s,t}$ in the two-pion DAs and the magnitudes of the weight factors $c_{\omega}$ and  $c_{\rho(1450)}$ can be  determined as
\begin{eqnarray}
	a_{p}^{0}=-0.7,\quad a_{p}^{s}=0.8,\quad a_{p}^{t}=-1.3,\quad \lvert c_{\rho(1450)} \rvert=0.11,\quad \lvert c_{\omega} \rvert=0.01.
\end{eqnarray}

As for the lineshap function, the Gounaris-Sakurai mass lineshape  $GS_{\rho}\left(s,m,\Gamma\right)$ is employed to describe the $\rho\left(770\right)$ resonance contribution~\cite{LHCb:2019sus}, and is given as
\begin{eqnarray} 
GS_{\rho}\left(s,m_{\rho},\Gamma_{\rho}\right)=\frac{m_{\rho}^{2}\left(1+D\Gamma_{\rho}/m_{\rho}\right)}{\left(m_{\rho}^{2}-s\right)+f\left(s\right)-im_{\rho}\Gamma(s)},
\end{eqnarray}
with the functions defined as
\begin{eqnarray} f\left(s\right)&=&\Gamma_{\rho}\frac{m_{\rho}^{2}}{p_{0}^{3}}\left[p_{1}^{2}\left[h\left(s\right)-h\left(m_{\rho}\right)\right]+\left(m_{\rho}^{2}-s\right)p_{0}^{2}\frac{\mathrm{d}h}{\mathrm{d}s}\bigg|_{\sqrt{s}=m_{\rho}}\right], \nonumber \\	h\left(s\right)&=&\frac{2}{\pi}\frac{p_{1}}{\sqrt{s}}\log\left(\frac{\sqrt{s}+2p_{1}}{2m_{\pi}}\right), \nonumber \\
\frac{\mathrm{d}h}{\mathrm{d}s}\bigg|_{\sqrt{s}=m_{\rho}}&=&h\left(m_{\rho}^2\right)\left[\left(8p_{0}^{2}\right)^{-1}-\left(2m_{\rho}^{2}\right)^{-1}\right]+\left(2\pi m_{\rho}^{2}\right)^{-1}, \nonumber \\
D&=&\frac{3}{\pi}\frac{m_{\pi}^{2}}{p_{0}^{2}}\log\left(\frac{m_{\rho}+2p_{0}}{2m_{\pi}}\right)+\frac{m_{\rho}}{2\pi p_{0}}-\frac{m_{\pi}^{2}m_{\rho}}{\pi p_{0}^{3}},
\end{eqnarray}
where the $p_{1}$ is the momentum of one daughter from $\rho(770)$ resonance, and is equal to $p_{0}$ when $s=m_{\rho}^2$. For the $\omega(782)$ and $\rho(1450)$ resonances, we follow Ref.~\cite{LHCb:2019sus} and adopt the relativistic Breit-Wigner (RBW) line shape \cite{ParticleDataGroup:2022pth}, which is given as
\begin{eqnarray}
RBW_R(s)=\frac{m_R^2}{m_R^2-s-im_R\Gamma(s)},
\label{RBW}
\end{eqnarray}
where $R=\omega(782)$ and $\rho(1450)$, and $m_R$ is their nominal mass. $\Gamma(s)$ is the mass-dependent width for the general case of a spin-$L$ resonance and is given  as
\begin{eqnarray}
\Gamma(s)=\Gamma_R\left(\frac{\mid\vec{q}\mid}{\mid\vec{q}_R\mid}\right)^{2L+1}\left(\frac{m_R}{\sqrt s}\right)X_L^2(\zeta),
\end{eqnarray}
where $\mid\vec{q}\mid$ is the momentum of one of $\pi$ in the $\pi\pi$ pair, and is equal to $\mid\vec{q}_R\mid$ when $s=m_R^2$. The values of $\Gamma_R$ and $m_R$ corresponding to the $\omega(782)$ and $\rho(1450)$ can be found in Ref. \cite{ParticleDataGroup:2022pth}. $X_L(\zeta)$ is the Blatt-Weisskopf angular momentum barrier factor \cite{Blatt:1952ije}
\begin{eqnarray}
&&L=0:\;\;X_L(\zeta)=1,\\
&&L=1:\;\;X_L(\zeta)=\sqrt{\frac{1+\zeta_0^2}{1+\zeta^2}},\\
&&L=2:\;\;X_L(\zeta)=\sqrt{\frac{9+3\zeta_0^2+\zeta_0^4}{9+3\zeta^2+\zeta^4}},
\end{eqnarray}
with $\zeta=r|\vec{q}|$ with $r$ being the effective meson radius, and $\zeta_0$ being the value of the $\zeta$ when $s=m_R^2$. It should be noted that in Refs.\cite{Wang:2016rlo, Li:2017obb, BaBar:2012bdw}, the authors adopted the Gounaris-Sakurai mass lineshap for describing the contribution of $\omega(782)$.

For the $D$-wave function of $\pi\pi$ pair, similar to the case of $ P$-wave $\pi\pi$ pair, only the longitudinal polarization state contributes to the decay amplitudes. The longitudinal component of $D$-wave function of $\pi\pi$ pair is similar to the  $P$-wave case, which is explicitly written as \cite{Zou:2020ool},
\begin{eqnarray} \Phi_{D}^{L}\left(z,\zeta,s\right)=\sqrt{\frac{2}{3}}\dfrac{1}{\sqrt{2N_{C}}}\left[\slashed{p}\phi_{D}^{0}\left(z,\zeta,s\right)+\omega\phi_{D}^{s}\left(z,\zeta,s\right)+\frac{\slashed{p}_{1}\slashed{p}_{2}-\slashed{p}_{2}\slashed{p}_{1}}{\sqrt{s}\left(2\zeta-1\right)}\phi_{D}^{t}\left(z,\zeta,s\right)\right]. \ \ \
\end{eqnarray}
The twist-2 and twist-3 DAs are
\begin{eqnarray}
	\phi_{D}^{0}\left(z,\zeta,s\right)&=&\frac{9F_{D}^{\parallel}\left(s\right)}{\sqrt{2N_{C}}}a_{D}z\left(1-z\right)\left(2z-1\right)P_{2}\left(2\zeta-1\right),\nonumber \\
	\phi_{D}^{s}\left(z,\zeta,s\right)&=&-\frac{9F_{D}^{\perp}\left(s\right)}{4\sqrt{2N_{C}}}a_{D}\left(1-6z+6z^{2}\right)P_{2}\left(2\zeta-1\right),\nonumber \\
	\phi_{D}^{t}\left(z,\zeta,s\right)&=&\frac{9F_{D}^{\perp}\left(s\right)}{4\sqrt{2N_{C}}}a_{D}\left(2z-1\right)\left(1-6z+6z^{2}\right)P_{2}\left(2\zeta-1\right),
\end{eqnarray}
where $P_{2}\left(2\zeta-1\right)=1-6\zeta+6\zeta^{2}$ is the legendre polynomial, and the Gegenbauer moment $a_D=1.1$ is determined based on the analysis in Ref.~\cite{LHCb:2019sus}. The $D$-wave time-like form factors $F_{D}^{\perp}$ and $F_{D}^{\parallel}$ also satisfy the relation $F_{D}^{\perp}/F_{D}^{\parallel}\approx f_{D}^{T}/f_{D}$, and $F_{D}^{\parallel}$ can also be described by the RBW lineshape shown in Eq.~\eqref{RBW}.

Using the theoretical framework introduced above, we then calculate the decay amplitudes of $B^{+}\to\pi^{+}\left(R\to\right)\pi^{+}\pi^{-}$ ($R=\rho(770), \omega(782), \rho(1450), f_2(1270)$) decays. Their explicit forms are given in the appendix A. With these obtained amplitudes, we can further evaluate the observables of the decay considered in this work. The differential branching fraction of the quasi-two-body $B$ meson decay is~\cite{El-Bennich:2009gqk}
\begin{eqnarray}
	\frac{\mathrm{d}^{2}\mathcal{B}}{\mathrm{d}\zeta\mathrm{d}\omega}=\tau_{B}\frac{\lvert\vec{p_{1}}\rvert\lvert\vec{p_{3}}\rvert}{32\pi^{3}M_{B}^{2}}\lvert\mathcal{A}\rvert^{2},
\end{eqnarray}
where
\begin{eqnarray}
	\lvert\vec{p_{1}}\rvert=\frac{\lambda^{1/2}\left(\sqrt s,m_{\pi},m_{\pi}\right)}{2\omega},\quad
	\lvert\vec{p_{3}}\rvert=\frac{\lambda^{1/2}\left(M_{B},m_{\pi},\sqrt s\right)}{2M_{B}},
\end{eqnarray}
with $\lambda\left(a,b,c\right)=\left[a^{2}-\left(b+c\right)^{2}\right]\left[a^{2}-\left(b-c\right)^{2}\right]$.

\section{Numerical results and discussion}
\begin{table}[t]
	\renewcommand\arraystretch{1.0}
	\caption{The values of input parameters. The decay constants of intermediate resonance are given in Ref.~\cite{Cheng:2010yd,Cheng:2011fk,Li:2000zb}, while other parameters are given in PDG~\cite{ParticleDataGroup:2022pth}.}  
	\centering
	\begin{tabular}{c|lll}
		\hline
		\hline
		\multirow{2}{*}{Masses (GeV)} & $m_{B}=5.279$\,,\quad$m_{\pi}=0.140$\,,\quad$m_{\rho_{770}}=0.775$\,,\quad$m_{\omega}=0.783$\,,\\
		& $m_{\rho_{1450}}=1.465$\,,\quad$m_{f_{2}}=1.276$\\
		\hline
		\multirow{2}{*}{Decay constants (MeV)} & $f_{B}=210\pm20$\,,\quad$f_{\rho_{770}}=216\pm3$\,,\quad$f_{\rho_{770}}^{T}=165\pm9$\,,\quad$f_{\omega}=187\pm5$\,,\\
		&$f_{\omega}^{T}=151\pm9$\,,\quad$f_{f_{2}}=102\pm6$\,,\quad$f_{f_{2}}^{T}=117\pm25$\\
		\hline
		Decay widths (MeV) & $\Gamma_{\rho_{770}}=149.1$\,,\quad$\Gamma_{\omega}=8.49$\,,\quad$\Gamma_{\rho_{1450}}=400$\,,\quad$\Gamma_{f_{2}}=186.7$\\
		\hline
		\multirow{2}{*}{Wolfenstein parameters}  &  $A=0.826^{+0.018}_{-0.015}$\,,\quad  $\lambda=0.22500\pm0.00067$\,,\\
		& $\bar{\rho}=0.159\pm0.010$\,,\quad $\bar{\eta}=0.348\pm0.010$\\
		\hline
		\hline
	\end{tabular}
\label{Tab.I}
\end{table}

\begin{table}[t]
	\renewcommand\arraystretch{1.0}
	\centering
	\caption {The $CP$-averaged branching ratios for $B^{+}\to\pi^{+}(R\to)\pi^{+}\pi^{-}$ ($R=\rho(770)$,\, $\omega(782)$, \,$\rho(1450)$, \,$f_2(1270)$ ) decays.  The numbers in the ``PQCD'' column correspond to our theoretical results  within PQCD approach, and the ones
in the ``Exp.'' column are the experimental date~\cite{LHCb:2019sus}. }  
	\begin{tabular}{c| c| c}
		\hline
		\hline
		Decay modes & PQCD & Exp.\\
		\hline
		$B^{+}\to\pi^{+}\left(\rho\left(770\right)\to\right)\pi^{+}\pi^{-}$ & $10.47_{-3.63-1.06-0.94}^{+4.79+1.42+1.11}\times10^{-6}$ & $8.44\pm0.09\pm0.06\pm0.38\times10^{-6}$\\
		\hline
		$B^{+}\to\pi^{+}\left(\omega\left(782\right)\to\right)\pi^{+}\pi^{-}$ & $7.83_{-3.56-1.53-0.41}^{+4.14+1.35+0.41}\times10^{-8}$ & $7.60\pm0.46\pm0.15\pm0.61\times10^{-8}$ \\
		\hline
		$B^{+}\to\pi^{+}\left(\rho\left(1450\right)\to\right)\pi^{+}\pi^{-}$ & $6.68_{-2.49-0.94-0.61}^{+3.00+1.18+0.68}\times10^{-7}$ & $7.90\pm0.46\pm0.30\pm2.89\times10^{-7}$\\
		\hline
		$B^{+}\to\pi^{+}\left(f_2(1270)\to\right)\pi^{+}\pi^{-}$ & $1.27_{-0.78-0.22-0.18}^{+0.69+0.17+0.15}\times10^{-6}$ & $1.37\pm0.05\pm0.11\pm0.21\times10^{-6}$\\
		\hline
		\hline
	\end{tabular}
\label{Tab.II}
\end{table}

\begin{table}[t]
	\renewcommand\arraystretch{1.0}
	\centering
	\caption{ The direct CP asymmetries parameters (in units of $\%$).  Other legends are the same as those of Table~\ref{Tab.II}. }
	\begin{tabular}{c| c| c}
		\hline
		\hline
		Decay modes & PQCD & Exp. \\
		\hline
		$B^{+}\to\pi^{+}\left(\rho\left(770\right)\to\right)\pi^{+}\pi^{-}$ & $1.5_{-2.2-2.9-0.0}^{+2.1+0.1+0.0}$ & $0.7\pm1.1\pm0.6\pm1.5$\\
		\hline
		$B^{+}\to\pi^{+}\left(\omega\left(782\right)\to\right)\pi^{+}\pi^{-}$ & $40.6_{-9.8-8.0-0.6}^{+10.8+0.9+0.8}$ & $-4.8\pm6.5\pm1.3\pm3.5$\\
		\hline
		$B^{+}\to\pi^{+}\left(\rho\left(1450\right)\to\right)\pi^{+}\pi^{-}$ & $-1.4_{-1.2-1.1-0.0}^{+1.2+0.3+0.0}$ & $-12.9\pm3.3\pm3.6\pm35.7$\\
		\hline
		$B^{+}\to\pi^{+}\left(f_2(1270)\to\right)\pi^{+}\pi^{-}$ & $73.6_{-13.0-19.4-1.7}^{+6.7+3.3+1.7}$ & $46.8\pm6.1\pm1.5\pm4.4$\\
		\hline
		\hline
	\end{tabular}
	\label{Tab.III}
\end{table}

The values of input parameters used in the numerical calculation, such as masses, decay constants, decay widths and the Wolfenstein parameters of CKM matrix, are summarized in Table~\ref{Tab.I}. Using these parameters, we present our numerical results of the $CP$ averaged branching fractions and the CP asymmetries, which are listed in Tables.~\ref{Tab.II} and \ref{Tab.III}, respectively. In the calculation, three kinds of theoretical uncertainties are evaluated as has been shown in the second columns of Tables.~\ref{Tab.II} and \ref{Tab.III}.  The first error is caused by the parameters in the distribution amplitudes of the $B$ meson, $\pi$ meson, and the $\pi\pi$ pair, such as the decay constants, shape parameter and the Gegenbauer moments. The second error is caused by the higher order radiative corrections and higher power corrections, which are evaluated by varying the  $\Lambda_{QCD}= \left(0.25\pm0.05\right)~{\rm GeV}$ and the factorization scale $t$ from $0.8t$ to $1.2t$. The last error arises from the uncertainties of the CKM matrix elements.

From Table~\ref{Tab.II}, it can be found that our theoretical results for the branching fractions of $B^{+}\to\pi^{+}\left(R\to\right)\pi^{+}\pi^{-}$ ($R=\rho(770), \omega(782), \rho(1450), f_2(1270)$) decays in the PQCD approach are in good consistence with the experimental data reported by the LHCb collaboration within errors. It should be noted that the theoretical errors are dominated by the nonperturbative parameters, such as the Gegenbauer moments in the wave function of the $\pi\pi$ pair. In fact, the theoretical description of the two-pion wavefunction is still in the stage of the modeling, the forms determined through the phenomenological models have to be used in the calculation, because the result based on the first principle is absent. As a result, the significant uncertainties caused by the hadronic inputs are unavoidable for now, more efforts are required for improving precision of the theoretical predictions.

It is well known to us that the $\omega(782)$ meson decaying directly to two pions is prohibited strictly due to the conservation of the $G$ parity, however this $G$-parity violating process can occur via $\rho-\omega$ mixing effect.  Some works indicate the potential importance of the $\rho-\omega$  mixing effect between the $\omega(782)$ and $\rho(770)$ resonances \cite{Wang:2015ula, Guo:2000uc, Cheng:2016shb}. In order to describe such effects, a model is proposed to  directly parameterize the interference between these two contributions~\cite{LHCb:2019sus}. In this work, such effect is also considered in a similar way via Eq.~\eqref{lineshape}. Because the contribution associated with  $\omega(782)$  is induced only by the $\omega-\rho$ mixing effect, the branching ratio of the $B^{+}\to\pi^{+} \left(\omega\left(782\right) \to\right) \pi^{+}\pi^{-}$ is expected to be far smaller than that of decay with intermediate resonance  $\rho(770)$ in principle. It can be clearly seen from the numerical results given in Table~\ref{Tab.II},  ${\cal B}(B^{+}\to\pi^{+}\left(\omega\to)\pi^{+}\pi^{-}\right)/{\cal B}(B^{+}\to\pi^{+}\left(\rho\to)\pi^{+}\pi^{-}\right) \sim {\cal O}(10^{-2})$.

In order to further test the framework of the quasi-two-body $B$ decays in PQCD approach and the wave functions of two-pion pair,  we also calculate the branching fractions of two-body $B^+\to \pi^+R$ decays by using the obtained results for the corresponding  quasi-two-body $B^{+}\to\pi^{+}\left(R\to\right)\pi^{+}\pi^{-}$ decays and the experimental data for $R\to\pi^{+}\pi^{-}$ decays. It is known that under the narrow-width-approximation (NWA). a branching fraction of the quasi-two-body decay can be decomposed as
\begin{eqnarray}
	\mathcal{B}\left(B^{+}\to\pi^{+}\left(R\to\right)\pi^{+}\pi^{-}\right)\simeq\mathcal{B}\left(B^{+}\to\pi^{+}R\right)\times\mathcal{B}\left(R\to\pi^{+}\pi^{-}\right),
\end{eqnarray}
The branching fractions of $R\to\pi^{+}\pi^{-}$~($R=\rho\,,\omega$ and $f_2$) decays have been measured with high precision, and the experimental results averaged by PDG are \cite{ParticleDataGroup:2022pth}
\begin{eqnarray}
\mathcal{B}\left(\rho\to\pi^{+}\pi^{-}\right)&\simeq & 100\%,\\
\mathcal{B}\left(\omega\to\pi^{+}\pi^{-}\right)&=&(1.53^{+0.11}_{-0.13})\%,\\
\mathcal{B}\left(f_{2}\to\pi\pi\right)&=&(84.2^{+2.9}_{-0.9})\%.
\label{eq:data2}
\end{eqnarray}
The experimental results for ${\cal B}(f_{2}\to\pi^+\pi^-)$ can be obtained by using the relation ${\cal B}(f_{2}\to\pi^+\pi^-)=2/3\times{\cal B}( f_{2}\to\pi\pi)$ required by the conservation of $G$ parity. Then, using these experimental data and our results for the corresponding  quasi-two-body $B^{+}\to\pi^{+}\left(R\to\right)\pi^{+}\pi^{-}$  decays listed in Table~\ref{Tab.II}, we can finally obtain
\begin{eqnarray}
	\mathcal{B}\left(B^{+}\to\pi^{+}\rho\right)&=&\frac{\mathcal{B}\left(B^{+}\to\pi^{+}\left(\rho\to\right)\pi^{+}\pi^{-}\right)}{\mathcal{B}\left(\rho\to\pi^{+}\pi^{-}\right)}=\left(10.47^{+0.00+7.32}_{-0.00-5.64}\right)\times10^{-6},\\
	\mathcal{B}\left(B^{+}\to\pi^{+}\omega\right)&=&\frac{\mathcal{B}\left(B^{+}\to\pi^{+}\left(\omega\to\right)\pi^{+}\pi^{-}\right)}{\mathcal{B}\left(\omega\to\pi^{+}\pi^{-}\right)}=\left(5.12^{+0.37+3.87}_{-0.43-4.29}\right)\times10^{-6},\\
	\mathcal{B}\left(B^{+}\to\pi^{+}f_{2}\right)&=&\frac{\mathcal{B}\left(B^{+}\to\pi^{+}\left(f_{2}\to\right)\pi^{+}\pi^{-}\right)}{2/3\times\mathcal{B}\left(f_{2}\to\pi\pi\right)}=\left(2.26^{+0.08+1.79}_{-0.04-2.09}\right)\times10^{-6},
\end{eqnarray}
where the first error corresponds to the experimental error  and the second one corresponds to  the theoretical error given in Table. ~\ref{Tab.II}. In comparison with the experimental results \cite{ParticleDataGroup:2022pth},
\begin{eqnarray}
	\mathcal{B}\left(B^{+}\to\pi^{+}\rho\right)&=&\left(8.3\pm1.2\right)\times10^{-6} , \\
	\mathcal{B}\left(B^{+}\to\pi^{+}\omega\right)&=&\left(6.9\pm0.5\right)\times10^{-6} , \\
	\mathcal{B}\left(B^{+}\to\pi^{+}f_{2}\right)&=&\left(2.2^{+0.7}_{-0.4}\right)\times10^{-6},
\end{eqnarray}
it can be found that our theoretical results are in agreement with experimental data  within the uncertainties.

At last, we turn to discuss the results of the direct $CP$ asymmetries in these decays, which are presented in Table.~\ref{Tab.III}. It can be found from the table that the direct $CP$ asymmetries of $B^+\to\pi^+ (\rho(770,1450)\to)\pi^+\pi^-$ decays are very small, which are consistent with the experimental data given by LHCb. It is known that the direct $CP$ asymmetry is mainly induced by the  interference between the tree and penguin contributions. However, the decays $B^+\to\pi^+ (\rho(770,1450)\to)\pi^+\pi^-$ are dominated by the tree contributions from $b\to du\bar{u}$ process with large Wilson coefficients, and the contributions of annihilation diagrams associated with penguin operators with $q=u$ and $d$ significantly are canceled by each other, which can be clearly seen from Eq.~\eqref{eq:A1}.  For the $B^+\to\pi^+ (f_2(1270)\to)\pi^+\pi^-$ decay, it has a large $CP$ asymmetry as large as $73.6\%$ due to the fact that the penguin contributions from chiral enhanced annihilation diagrams no longer cancel each other out. Our theoretical results agree with the analysis of LHCb collaboration within errors.  Similarly, a relatively large  $CP$ asymmetry is expected also by PQCD in the $B^+\to\pi^+ (\omega\to)\pi^+\pi^-$ decay, which could be tested in LHCb and Belle-II experiments in future.

\section{Summary}
In this paper, motivated by the analysis of  $B^{+}\to\pi^{+}\pi^{-}\pi^{+}$ decay made by LHCb collaboration, we investigated the  $B^{+}\to\pi^{+}(R\to)\pi^{+}\pi^{-}$ decays with the P-wave $\rho(770)$, $\omega(782)$, $\rho(1450)$ and D-wave $f_2(1270)$ intermediate resonant states within the framework of PQCD approach, in which the effect of $\rho-\omega$ mixing are considered. The branching fractions and direct CP asymmetries of these decays are calculated, our numerical results are summarized in Tables~\ref{Tab.II} and \ref{Tab.III}.  Our results for ${\cal B}(B^{+}\to\pi^{+}(R\to)\pi^{+}\pi^{-})$~($R=\rho(770)$, $\omega(782)$, $\rho(1450)$, $f_2(1270)$), as well as the extracted  ${\cal B}(B^{+}\to\pi^{+}(\rho(770), \omega(782), f_2(1270) ))$, are consistent with the experimental data within errors. The theoretical results for the direct $CP$ asymmetries of tree dominated $B^+\to\pi^+ (\rho(770/1450)\to)\pi^+\pi^-$ decays are very small,  while the ones of tree and penguin dominated $B^+\to\pi^+ (\omega/f_2\to)\pi^+\pi^-$ are relatively large. Most of them agree well with the LHCb data except for $B^+\to\pi^+ (\omega\to)\pi^+\pi^-$ decay, which occurs only via $\rho-\omega$ mixing effect. More theoretical and experimental efforts are needed for clarifying the underling mechanism of such special decay mode.    


\section*{Acknowledgment}
 This work is supported by the National Natural Science Foundation of China (Grant Nos. 12275067, 12305101, 11875122, 12375089), Natural Science Foundation of Shandong province (Grant no. ZR2022MA035 and ZR2022ZD26),   Natural Science Foundation of Henan Province (Grant No. 225200810030), and   Excellent Youth Foundation of Henan Province (Grant No. 212300410010).  

\appendix
{\centering\section*{Appendix A:Decay amplitudes}}
The decays amplitudes of decay modes studied  in this work are given as
\setcounter{equation}{0}
\renewcommand{\theequation}{A.\arabic{equation}}
\begin{align}
\mathcal{A}\left(B^{+}\to\pi^{+}\left(\rho_{770}/\rho_{1450}\to\right)\pi^{+}\pi^{-}\right)=& \frac{G_{F}}{2}\Big\{V_{ub}^{*}V_{ud}\big[C_{1}(M_{VP,P}^{LL}+W_{VP,P}^{LL}-W_{VP,V}^{LL})+C_{2}M_{VP,V}^{LL}\nonumber\\
&+(\frac{C_{1}}{3}+C_{2})(F_{VP,P}^{LL}+A_{VP,P}^{LL}-A_{VP,V}^{LL})+(C_{1}+\frac{C_{2}}{3})F_{VP,V}^{LL}\big]\nonumber\\
&-V_{tb}^{*}V_{td}\big[(C_{3}+C_{9})(M_{VP,P}^{LL}+W_{VP,P}^{LL}-W_{VP,V}^{LL})+(C_{5}+C_{7})\nonumber\\
&(M_{VP,P}^{LR}+W_{VP,P}^{LR}-W_{VP,V}^{LR})+(\frac{C_{3}}{3}+C_{4}+\frac{C_{9}}{3}+C_{10})(F_{VP,P}^{LL}\nonumber\\
&+A_{VP,P}^{LL}-A_{VP,V}^{LL})+(\frac{C_{5}}{3}+C_{6}+\frac{C_{7}}{3}+C_{8})(F_{VP,P}^{SP}+A_{VP,P}^{SP}\nonumber\\
&-A_{VP,V}^{SP})+(-\frac{C_{3}}{3}-C_{4}+\frac{3}{2}C_{7}+\frac{1}{2}C_{8}+\frac{5}{3}C_{9}+C_{10})F_{VP,V}^{LL}\nonumber\\
&+(-C_{3}+\frac{1}{2}C_{9}+\frac{3}{2}C_{10})M_{VP,V}^{LL}+(-C_{5}+\frac{1}{2}C_{7})M_{VP,V}^{LR}\nonumber\\
&+\frac{3}{2}C_{8}M_{VP,V}^{SP}\big]\Big\}.	
\label{eq:A1}	
\end{align}
\begin{align}
	\mathcal{A}\left(B^{+}\to\pi^{+}\left(\omega\left(782\right)\to\right)\pi^{+}\pi^{-}\right)=& \frac{G_{F}}{2}\Big\{V_{ub}^{*}V_{ud}\big[C_{1}(M_{VP,P}^{LL}+W_{VP,P}^{LL}+W_{VP,V}^{LL})+C_{2}M_{VP,V}^{LL}\nonumber\\
	&+(\frac{C_{1}}{3}+C_{2})(F_{VP,P}^{LL}+A_{VP,P}^{LL}+A_{VP,V}^{LL})+(C_{1}+\frac{C_{2}}{3})F_{VP,V}^{LL}\big]\nonumber\\
	&-V_{tb}^{*}V_{td}\big[(C_{3}+C_{9})(M_{VP,P}^{LL}+W_{VP,P}^{LL}+W_{VP,V}^{LL})+(C_{5}+C_{7})\nonumber\\
	&(M_{VP,P}^{LR}+W_{VP,P}^{LR}+W_{VP,V}^{LR})+(\frac{C_{3}}{3}+C_{4}+\frac{C_{9}}{3}+C_{10})(F_{VP,P}^{LL}\nonumber\\
	&+A_{VP,P}^{LL}+A_{VP,V}^{LL})+(\frac{C_{5}}{3}+C_{6}+\frac{C_{7}}{3}+C_{8})(F_{VP,P}^{SP}+A_{VP,P}^{SP}\nonumber\\
	&+A_{VP,V}^{SP})+(\frac{7}{3}C_{3}+\frac{5}{3}C_{4}+2C_{5}+\frac{2}{3}C_{6}+\frac{1}{2}C_{7}+\frac{1}{2}C_{8}+\frac{1}{3}C_{9}\nonumber\\
	&-\frac{1}{3}C_{10})F_{VP,V}^{LL}+(C_{3}+2C_{4}-\frac{1}{2}C_{9}+\frac{1}{2}C_{10})M_{VP,V}^{LL}\nonumber\\
	&+(C_{5}-\frac{1}{2}C_{7})M_{VP,V}^{LR}+(2C_{6}+\frac{1}{2}C_{8})M_{VP,V}^{SP}\big]\Big\}.		
\end{align}
The contributions of the $n\bar{n}=\frac{1}{\sqrt{2}}\left(u\bar{u}+d\bar{d}\right)$ and the ($s\bar{s}$) quark components to the decay amplitude of the $B^{+}\to\pi^{+}\left(f_{2}\left(1270\right)\to\right)\pi^{+}\pi^{-}$ are written as
\begin{align}
\mathcal{A}_{n\bar{n}}\left(B^{+}\to\pi^{+}\left(f_{2}\left(1270\right)\to\right)\pi^{+}\pi^{-}\right)=& \frac{G_{F}}{2}\Big\{V_{ub}^{*}V_{ud}\big[C_{1}(M_{TP,P}^{LL}+W_{TP,P}^{LL}+W_{TP,T}^{LL})+C_{2}M_{TP,T}^{LL}\nonumber\\
	&+(\frac{C_{1}}{3}+C_{2})(F_{TP,P}^{LL}+A_{TP,P}^{LL}+A_{TP,T}^{LL})\big]\nonumber\\
	&-V_{tb}^{*}V_{td}\big[(C_{3}+C_{9})(M_{TP,P}^{LL}+W_{TP,P}^{LL}+W_{TP,T}^{LL})+(C_{5}+C_{7})\nonumber\\
	&(M_{TP,P}^{LR}+W_{TP,P}^{LR}+W_{TP,T}^{LR})+(\frac{C_{3}}{3}+C_{4}+\frac{C_{9}}{3}+C_{10})(F_{TP,P}^{LL}\nonumber\\
	&+A_{TP,P}^{LL}+A_{TP,T}^{LL})+(\frac{C_{5}}{3}+C_{6}+\frac{C_{7}}{3}+C_{8})(F_{TP,P}^{SP}+A_{TP,P}^{SP}\nonumber\\
	&+A_{TP,T}^{SP})+(C_{3}+2C_{4}-\frac{1}{2}C_{9}+\frac{1}{2}C_{10})M_{TP,T}^{LL}\nonumber\\
	&+(C_{5}-\frac{1}{2}C_{7})M_{TP,T}^{LR}+(2C_{6}+\frac{1}{2}C_{8})M_{TP,T}^{SP})\big]\Big\}\,,	\\
\mathcal{A}_{s\bar{s}}\left(B^{+}\to\pi^{+}\left(f_{2}\left(1270\right)\to\right)\pi^{+}\pi^{-}\right)=&-\frac{G_{F}}{\sqrt{2}}V_{tb}^{*}V_{td}\big[(C_{4}-\frac{1}{2}C_{10})M_{TP,T}^{LL}\nonumber\\
	&+(C_{6}-\frac{1}{2}C_{8})M_{TP,T}^{SP})\big]	\,,	
\end{align}
 respectively, where $G_{F}$ is the Fermi constant, $V_{ij}$ are the Cabibbo-Kobayashi-Maskawa matrix elements. In above formula,  the functions $F_{VP,P}^{LL},M_{VP,V}^{LR},A_{TP,P}^{SP}$, $W_{TP,T}^{LL},....$ are the individual decay amplitudes corresponding to different currents and diagrams;  $F$, $M$, $A$ and $W$ correspond to the factorable emission diagram, the non-factorable emission diagram, the factorable annihilation diagram, the non-factorable annihilation diagram, respectively; the subscript $P$, $V$ and $T$ denote the pseudoscalar, vector and tensor states, respectively; The superscripts $LL$,$LR$,and $SP$ denote $(V-A)(V-A)$, $(V-A)(V+A)$ and $(S-P)(S+P)$ currents, respectively.  The explicit expressions of these functions can be found in the Ref.~\cite{Li:2021cnd}. 


\end{document}